\documentclass[conference]{IEEEtran}
\IEEEoverridecommandlockouts
\usepackage{amsmath,amssymb,amsfonts}
\usepackage{graphicx}
\graphicspath{{./}{figure/}}
\usepackage{booktabs}
\usepackage{siunitx}
\usepackage{tabularx,multirow,makecell}
\usepackage{longtable}
\usepackage{colortbl}
\usepackage[table]{xcolor}
\usepackage{pdflscape}
\usepackage{float}
\usepackage{braket}
\usepackage{pifont}
\usepackage{comment}
\usepackage{balance}
\usepackage[utf8]{inputenc}   
\usepackage[T1]{fontenc}      
\usepackage{tikz}
\usetikzlibrary{arrows.meta,positioning,calc}
\usetikzlibrary{arrows.meta,positioning}
\usepackage{pgfplots}
\pgfplotsset{compat=1.18}
\usetikzlibrary{babel,plotmarks,positioning,arrows.meta,fit,shapes.geometric}
\tikzset{>=Latex}
\usepackage{cuted}
\usepackage{capt-of}
\usepackage[ruled,vlined,noend]{algorithm2e}
\SetKw{Break}{break} 
\usepackage{xcolor}

\SetCommentSty{mycommfont}
\SetAlFnt{\small}
\SetAlCapFnt{\small}
\SetAlCapNameFnt{\small}
\SetKwInOut{KwIn}{Input}
\SetKwInOut{KwOut}{Output}
\SetKw{KwRet}{Return}

\usepackage{cite}
\usepackage[hidelinks]{hyperref}
\usepackage[nameinlink,capitalise,noabbrev]{cleveref}
\usepackage[section]{placeins}
\crefname{algocf}{Algorithm}{Algorithms}
\Crefname{algocf}{Algorithm}{Algorithms}

\newcolumntype{Y}{>{\centering\arraybackslash}X}

\begin{document}

\title{Security Evaluation of Quantum Circuit Split Compilation under an Oracle-Guided Attack}
\author{
  \IEEEauthorblockN{Hongyu Zhang and Yuntao Liu}
  \IEEEauthorblockA{Department of Electrical and Computer Engineering, Lehigh University, PA, USA}
  \IEEEauthorblockA{\{hoz324, yule24\}@lehigh.edu}
}
\maketitle



\begin{abstract}

Quantum circuits are the fundamental representation of quantum algorithms and constitute valuable intellectual property (IP). Multiple quantum circuit obfuscation (QCO) techniques have been proposed in prior research to protect quantum circuit IP against malicious compilers. However, there has not been a thorough security evaluation of these schemes. In this work, we investigate the resilience of split compilation against an oracle-guided attack. Split compilation is one of the most studied QCO techniques, where the circuit to be compiled is split into two disjoint partitions. Each split circuit is known to the compiler, but the interconnections between them are hidden. We propose an oracle-guided security evaluation framework in which candidate connections are systematically tested against input-output observations, with iteratively pruned inconsistent mappings. This hierarchical matching process exploits the reversibility of quantum gates and reduces the search space compared to brute-force enumeration. Experimental evaluation in the RevLib benchmark suite shows that only a small number of I/O pairs are sufficient to recover the correct inter-split connections and reconstruct the entire circuits. Our study marks the first thorough security evaluations in quantum IP protection and highlights the necessity of such evaluations in the development of new protection schemes.

\end{abstract}

\maketitle

\section{Introduction}

The advancement of quantum computers ushers in a new era of computational power and methodologies. 
Quantum algorithms are implemented through single operations, called quantum gates, while gate sequences form quantum circuits, the hardware-executable representation of quantum programs. Cloud platforms such as IBM Quantum, Amazon Braket, and Microsoft Azure \cite{chow2021ibm, gonzalez2021cloud, prateek2023quantum} provide extensive access to quantum hardware, while compilers such asskit \cite{qiskit2024} map high-level circuit descriptions to device-specific instructions.
Despite these advances, quantum circuit design remains resource-intensive and requires substantial intellectual capital. Consequently, quantum circuit designs are increasingly recognized as valuable intellectual property (IP)~\cite{aboy2022mapping}, analogous to integrated circuits (ICs) in classical computing. Recent research has explored quantum-specific defense mechanisms—including split compilation~\cite{saki2021split, wang2025tetrislock}, insertion of randomized reversible subcircuits~\cite{das2023randomized, suresh2021short}, quantum logic locking~\cite{topaloglu2023quantum, liu2025loq}, and phase-level obfuscation techniques~\cite{rehman2025opaque} to protect against threats to designs posed by untrusted compilers and cloud environments.

Split compilation~\cite{saki2021split, wang2025tetrislock}, inserting random reversible circuits \cite{das2023randomized, suresh2021short}, quantum logic locking \cite{topaloglu2023quantum, liu2025loq}, and phase obfuscation \cite{rehman2025opaque} to protect against untrusted compilers capable of IP theft or Trojan insertion~\cite{das2023trojannet, john2025stealthy}.

These efforts assume that hiding or fragmenting circuit structure is sufficient to prevent adversaries from recovering original designs.
However, these works did not consider any sophisticated adversaries beyond brute-force attacks.
However, a significant gap remains: the problem of reconstructing circuits when their partitioned forms are exposed, but interconnections remain hidden. 
For example, the work on split compilation \cite{saki2021split, wang2025tetrislock} assumed that adversaries needed to exhaustively try all the permutations of qubit mappings between the two split circuits. This oversimplified assumption overlooked the structural and functional information that exists in the split circuits that can be leveraged to reduce the number of I/O queries needed to reconstruct the qubit mapping in the presence of a black-box oracle.

We propose a reverse engineering framework for quantum circuits protected by partitioned compilation. Attackers who know both partitioned circuits but lack knowledge of the qubit mapping can determine the circuit correspondence by selecting inputs and measuring outputs.

We employ a hierarchical reconstruction algorithm to process split-compiled circuits.
The process first stitches single-wire traces, then progressively expands to dual-layer, triple-layer, and ultimately complete boundary replacement, that is, the entire circuit, until parsed.
At the reconstruction level $k$ ($k=1,2,\ldots$), we generate candidate $k$ wire interconnect schemes while maintaining single-wire consistency; The surviving set $\mathcal{H}_k$ undergoes iterative validation through queries to a finite budget input-output prediction oracle, eliminating any hypotheses that fail input-output consistency checks.
Using gate reversibility and systematic block elimination mechanisms, this algorithm significantly reduces search space compared to brute-force methods that exhaustively explore all boundary permutations.
The main contributions of this paper are as follows.
\begin{itemize}
    \item \textbf{Problem formulation and query efficiency.}
  We transform the reverse engineering problem of split-compiled circuits into \emph{connection restoration} rather than an exhaustive permutation search. This reduces the number of required oracle queries under ideal conditions.

  \item \textbf{Hierarchical reconstruction.}
  We employ a layered approach, starting with single-wire consistency within each partitioned unit and progressively expanding to pairwise and multi-wire cross-partition hypotheses until the entire circuit is resolved. At each level, the candidate interconnections are validated against the observed input-output (\(x,y\)) behavior of the original circuit.

  \item \textbf{Security evaluation.}
  To our knowledge, this is the first systematic security evaluation of quantum \emph{partitioned (split) compilation} as an intellectual property protection technique, conducted on standard benchmarks under controlled query budgets.
\end{itemize}

Our proposed framework reconstructs the implicit interconnections between partitioned layers. This process first connects single paths within each partitioned layer, then progressively expands to establish paired and multipath connections between two partitioned layers, ultimately converging into a complete circuit. At each stage, candidate connections are validated using all input-output data across the circuit while eliminating inconsistent mapping relationships. Using the reversibility of quantum gates and employing a systematic block pruning strategy, this method significantly reduces the search space compared to brute-force enumeration.  

Experiments on the \textit{RevLib} benchmark demonstrate that only a small number of input-output pairs are required to recover the correct inter-segment connections. This approach substantially improves the efficiency of circuit reconstruction.

This research demonstrates the feasibility of reconstructing hidden interconnections between circuit segments and highlights the necessity of strengthening quantum intellectual property protection defense mechanisms in practical deployments.

\section{Background \& Related Work}

\subsection{Quantum Computing Fundamentals}
Quantum computing is one of the top five priority areas identified by the National Science Foundation for future investment and has already demonstrated clear advantages over classical computing in specific domains, marking significant milestones in computational science. One of the most notable achievements is Google's 2024 demonstration of quantum supremacy using its 105-qubit Willow processor  \cite{neven2024willow}. Willow completed a random circuit sampling task in under five minutes, while the same task would take classical supercomputers about 10 septillion years. This shows that quantum devices have shown the ability to outperform classical systems in computational tasks, including practical advantages in unstructured search, period finding, optimization, and machine learning.

Many quantum algorithms promise asymptotic speedups over their best-known classical counterparts, ranging from polynomial to exponential improvements. For example, the Quantum Approximate Optimization Algorithm (QAOA) targets combinatorial optimization with potential polynomial advantages \cite{Guerreschi2019, MontanezBarrera2025}, while the Variational Quantum Eigensolver (VQE) has shown promise in quantum chemistry applications \cite{peruzzo2014variational, kandala2017hardware}. At the other extreme, Shor’s algorithm achieves an exponential speedup for factoring and discrete logarithms using the Quantum Fourier Transform (QFT) \cite{Shor1994, ProosZalka2003}, and Grover’s algorithm provides a quadratic advantage for unstructured search \cite{grover1996fast}. 

Quantum computing relies on quantum bits, a.k.a. qubits, which differ from classical bits in that they can simultaneously exist in a superposition state of the ground states $\ket{0}$ and $\ket{1}$. We utilize quantum gates to apply unitary transformations to quantum bits, enabling their states to evolve reversibly. Quantum circuits are constructed by sequencing quantum gates to implement algorithms. Execution requires a quantum compiler to translate high-level circuit descriptions into hardware-specific operations, map logical qubits to physical qubits, and insert exchange operations when necessary \cite{chow2021ibm, gonzalez2021cloud, prateek2023quantum, qiskit2024}.

The reversibility of quantum gates is a crucial aspect of quantum computing, signifying that quantum processes can be counteracted by performing the operations in the opposite sequence. In contrast to classical logic gates, which can discard information (such as an AND gate that condenses two input bits into a single output bit), quantum gates are naturally reversible. This implies that their operations can be reversed to retrieve the system's initial state. This reversible feature originates from the unitary characteristics of quantum gates. For instance, if a quantum circuit executes gate $U$ followed by gate $V$, reversing the process requires applying $V^\dagger$ first and then $U^\dagger$. This property is essential for deriving the input state we use in oracle queries in our work, as we will see in Section \ref{ssec:attack} of this paper.

\subsection{Quantum Circuit IP Protection}
Similarly to classical integrated circuits, quantum circuits are recognized as valuable intellectual property (IP)\cite{aboy2022mapping}. In the setting of untrusted compilers and cloud execution, prior work is grouped into three broad lines. First, obfuscation alters the observable structure while preserving semantics: for example, inserting identity subcircuits (\(UU^\dagger\), where $U$ is any quantum gate), randomly permuting and renaming qubit wires, applying local Clifford conjugations to change gate appearance, or performing operator-level template rewrites, thus reducing structural similarity and gate-level readability at modest overheads in gate count and depth\cite{broadbent2014quantum, alagic2015circuit, alwen2015limited}. Second, quantum circuit locking introduces \(k\) key bits and controlled operations (e.g., controlled phases/rotations) such that only the unique correct key restores the target functionality; any incorrect key induces measurable deviations in the output statistics, yielding a key space of size \(2^{k}\). Locking locations are typically placed along high-impact cones to maximize error propagation while respecting synthesis and hardware-noise constraints\cite{xie2016sat, subramanyan2015evaluating}. Third, split compilation partitions a circuit along one or more cuts into multiple subcircuits (splits) so that no single compiler instance observes the entire design\cite{saki2021split, wang2025tetrislock}. Security claims here rest on combinatorial ambiguity at partition boundaries: exposing \(m\) boundary lines implies \(m!\) cross-partition permutations; with multiple independent cuts of sizes \(m_1,\ldots,m_R\), where $R$ is the number of cuts, the candidate space scales as \(\prod_{r=1}^{R} m_r!\). Variants further leverage asymmetric splits and interlocking patterns to enlarge the candidate space and suppress straightforward alignment\cite{saki2021split, wang2025tetrislock}. Collectively, these approaches provide complementary protection against IP threats from the compiler. Obfuscation lowers structural legibility, locking raises functional key complexity, and split compilation injects factorial-scale search at boundaries, with moderate increase in gates/depth and compilation constraints. 


\subsection{Reverse Engineering Gap}
Existing literature in QCO generally lacks operational and reproducible external verification frameworks on attack resilience. This absence prevents evaluating the effectiveness of these protective measures from an attacker's perspective, resulting in a lack of unified, comparable empirical benchmarks for their “effectiveness.” Therefore, we aim to select a method to verify whether the models and methods proposed by other scholars in QCO are as effective as claimed.
In classical integrated circuit obfuscation, key recovery attacks have long been regarded as a primary security threat \cite{subramanyan2015evaluating, rajendran2013split, xie20162}. Many of these attacks are oracle-guided, i.e., attackers validate potential keys after against known input and output pairs. 
Our work presents the first oracle-guide attack on QCO and enables attackers to efficiently recover the correct quantum bit mapping of quantum circuit split compilation with only a minimal number of input-output queries. 
Our research methodology serves both as an attacker-side mapping recovery algorithm and as an independent evaluation tool to verify whether other researchers' theoretical security claims align with their actual capabilities. Thus, our approach fills a gap in the quantum cryptography field regarding the study of cryptographic security reliability.


\section{Threat Models}

We investigate methods for exploiting recovery attacks to connect circuits within a partitioned compilation environment. The target circuit is divided into two disjoint parts (Split 1 and Split 2). The gate sequences within each split are known, but the interconnections between the partitions remain unknown.
Additionally, the attacker has oracle access to a working quantum circuit, i.e., they can set the input state and measure the output state of a correctly combined circuit, but cannot access the internal details of the oracle. 
Furthermore, we assume that the attacker knows the qubit order in Split 1. This is because in a deployed circuit, the context in which the oracle is used may hint at the application for which the quantum circuit is intended. Split 1 contains the start of the circuit and probably has the encoding of the problem, from which the attacker may infer the order of the qubit in Split 1.
As stated earlier, the attacker's goal is to recover the qubit mapping between the splits, and hence the whole quantum circuit, with the minimum number of input-output (I/O) queries.


\section{Connection Recovery under Split Compilation}


\subsection{Problem Setup and Overview}
\begin{figure*}[thb]
  \centering
  \includegraphics[width=\textwidth]{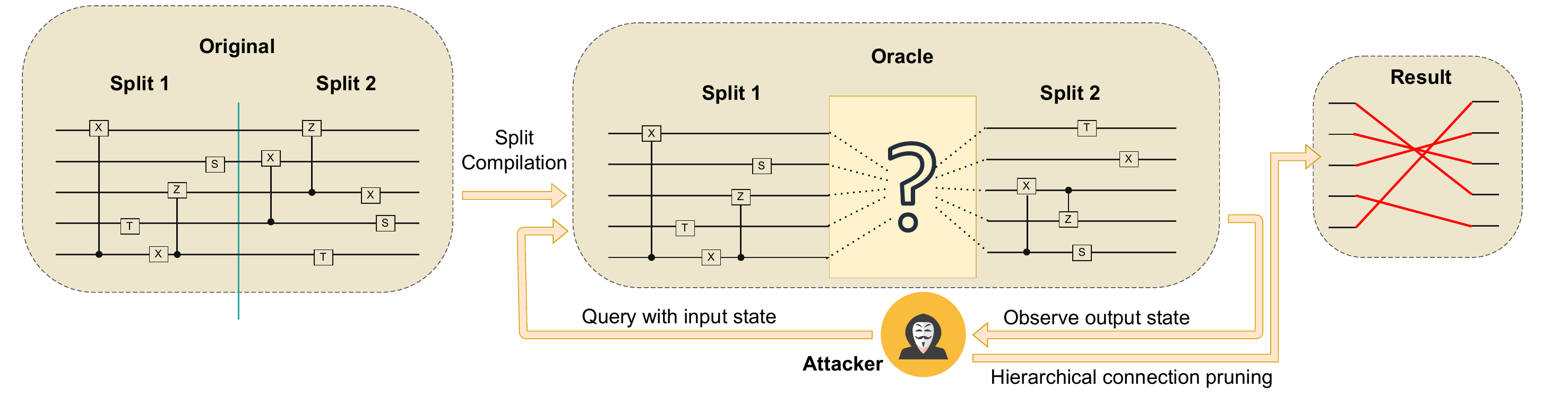}
  \caption{Flow of the oracle-guided attack model on split compilation. The original circuit is split into Splits 1 and 2 that are compiled separately. The attacker seeks to reverse-engineer the hidden qubit connections through input-output oracle queries on the deployed quantum circuit.}
  \label{fig:process}
\end{figure*}

Our attack targets split compilation scenarios, where the original circuit is divided into two partitions Split 1 and Split 2), resulting in two parts whose connection is unknown. 
Previous research on the compilation of quantum circuit splits \cite{saki2021split, wang2025tetrislock} assumed that the attacker must perform an exhaustive search to restore the qubit connections between the split circuits. However, security evaluation of split manufacturing and routing-based obfuscation of classical circuits suggests that, with appropriate formulations, the solution space can be pruned much more efficiently than a brute-force attack \cite{xie2016security, subramanyan2015evaluating, mcdaniel2024removal}. Therefore, we ask the natural question:
Can an attacker on quantum circuit split compilation recover the hidden interconnections between the two known partitions? 
We assume that the attacker knows: (1) the number of qubits \(m\) in the target circuit, (2) the complete gate sequences within \textit{Split 1} and \textit{Split 2}. Furthermore, the attacker has oracle access to a deployed quantum circuit, meaning that they can set the input state and measure the output state of the circuit but cannot see the circuit details.
The objective is to determine the interconnection mapping between the two split regions such that the reconstructed circuit \(C^\star\) matches the observed I/O behavior.
Figure~\ref{fig:process} outlines the overall flow of our connection restoration framework.

\subsection{Hierarchical Connection Recovery Method} 
\label{ssec:attack}
The theoretical basis of our method lies in leveraging the reversibility of quantum gates and circuits and progressively eliminating inconsistencies through hierarchical matching from simple to complex interconnections. Given the inputs and outputs of two splits, we validate candidate solutions based on observed input-output pairs. 
Our goal is to restore the hidden connections between Split 1 and Split 2, ensuring that the reconstructed circuit is functionally equivalent to the original circuit. When all inconsistent possible circuits are eliminated and only a single (or equivalence class) connection remains, our method is declared complete.


\subsubsection{Separating Qubits in Split 2 using Entanglement Relations}
We define a ``qubit block'' as a set of entangling qubits. In this way, we can map the qubits in each qubit block to the qubits in Split 1 separately. 
We use $B^b_i$ to indicate the $i$-th qubit block with $b$ qubits. Let $\mathcal{B}$ be the set of all qubit blocks.

\subsubsection{Block-Level Qubit Mapping}
To take advantage of the separability between qubit blocks and reduce the oracle queries needed, we first find the set of qubits in Split 1 corresponding to each qubit block in Split 2.
To this end, we generate input states to query the oracle in order to sensitize the qubits in Split 1 that are presumed to be connected to the block of interest in Split 2.
For each block $B^b_i$, we use $A^b_i$ to denote the set of qubits in Split 1 that is connected to it. In this step, we only need to identify the collection of qubits in Split 1 that maps to $B^b_i$. We will find the qubit-to-qubit mapping in the next step.

Among the qubits in Split 1 that are not assigned to any block in Split 2, we enumerate the combination of $b$ qubits and test if they for the right $A^b_{i}$. The test procedure is as follows.
Let us use $\ket{\Phi}=\ket{\varphi^1_0, \varphi^1_1, \ldots, \varphi^1_{m-1}}$ to denote the output state of the $j$-th qubit in Split 1, where $m$ is the total number of qutits. 
To sensitize the $b$ qubits in the currently tested $A^b_{i}$, we choose two target output states $\ket{\Phi}$ and $\ket{\Phi'}$, among which the qubits within $A^b_{i}$ have different output states, and all other qubits have the same state. The corresponding input states can be derived using the reversibility of Split 1.
Both inputs are applied to the oracle and the output states are observed, from which we can observe whether the current $A^b_{i}$ is the correct one. If so, any qubit outside $B^b_i$ should have the same output states for both inputs. If this is not the case, we attempt more combinations for $A^b_{i}$ until we find a match. We then mark the qubits in $A^b_{i}$ as assigned and move on the next qubit block in Split 2.
We map the qubit blocks in ascending order of the number of qubits $b$ in order to minimize the number of oracle queries needed.

\subsubsection{Pruning Qubit Mapping Permutations}
Once we identify the Split 1 qubits corresponding to a qubit block in Split 2, we need to find their qubit-to-qubit mapping in order to fully reverse engineer the quantum circuit.
To this end, we enumerate each qubit permutation of $B^b_i$ until the simulated output matches the correct one observed from the oracle.

\subsubsection{Accounting for Noise in Oracle Measurement}
Since the oracle is deployed on quantum hardware and the current generation of quantum hardware is noisy, we use an error threshold $\epsilon$ to differentiate between noisy results of the correct qubit mapping and those of incorrect qubit mappings. Let $\ket{O}$ and $\ket{O}'$ be the qubit states that are supposed to match. When their fidelity $F(\ket{O},\ket{O}') = |\braket{O|O'}|^2 \ge 1-\epsilon$, we consider the two states are the same. Otherwise, we consider they mismatch due to incorrect qubit mapping between Split 1 and Split 2.

\begin{algorithm}[thb]
  \caption{Hierarchical Connection Recovery}
  \DontPrintSemicolon
  \SetKwInOut{KwIn}{Input}
  \SetKwInOut{KwOut}{Output}
  \KwIn{Oracle $\mathcal{O}$; number of qubits $m$, hierarchical structure of Split 1  ($\mathcal{S}_1$) and Split 2 ($\mathcal{S}_2$), error threshold $\epsilon$.}
  \KwOut{Oracle query count $t$ or timeout (TO)}
    $t \gets 0$ \\
  \For{$b$ from 1 to $m$}{
    \ForEach{$B^b_i$}{
    \ForEach{candidate $A^b_{i}$}{
        \tcc{Sensitizing qubit block $B^b_i$}
        Generate $\ket{\Phi}=\ket{\varphi^1_0, \varphi^1_1, \ldots, \varphi^1_{m-1}}$ and $\ket{\Phi'}=\ket{\varphi'^1_0, \varphi'^1_1, \ldots, \varphi'^1_{m-1}}$ such $\ket{\phi_j}=\ket{\phi'_j}$ for any qubit $q_j \notin A^b_i$, and $\ket{\phi_j}\ne\ket{\phi'_j}$ for any qubit $q_j \in A^b_i$\\
        Calculate input state $\ket{I}=\mathbf{S}_1^{-1}(\ket{\Phi}),\ \ket{I'}=\mathbf{S}_1^{-1}(\ket{\Phi'}).$ \\
        Query oracle to obtain correct output $\ket{O}=\mathcal{O}(\ket{\Phi}),\ \ket{O'}=\mathcal{O}(\ket{\Phi'}).$\\
        $t \gets t+2$ \\
        Remove qubits inside $B^b_i$ from $\ket{O}$ and $\ket{O'}$, then heck whether $F(\ket{O},\ket{O'} \ge 1-\epsilon$. \\
        \uIf{match}{
            \tcc{Qubit permutation for $B^b_i$}
            \ForEach{Qubit permutation of $B^b_i$}{
                Concatenate $B^b_i$ with Split 1 to build simulation oracle $\mathcal{O}'$.\\
                Randomly sample input state $\ket{I}''$
                $t \gets t+1$ \\
                \uIf{$F(\mathcal{O}(\ket{I}''),\mathcal{O}'(\ket{I}''))\ge 1-\epsilon$ for qubits in $B^b_i$}{
                    Correct qubit mapping of $B^b_i$ is found.                    
                }
            }
    }
    }
    \If{Time elapsed > time limit}{
      \KwRet{TO} 
    }
  }
  }
\end{algorithm}

In summary, for each qubit block, we need to record a set of data to facilitate a final comparison.
To minimize attack complexity, we apply single-qubit pruning, double-qubit pruning, and multiqubit pruning sequentially. When inference stalls or match is not fixed, request new input-output pairs from , update the matched qubits pairs, frozen them and continue. The process terminates when we find that unique or indistinguishable equivalence classes exist under query I/O. 


\subsection{Advantages Over Brute-Force Search}
Previous studies assumed that attackers cannot infer connections between two splits unless through brute-force enumeration. The brute-force enumeration requires examining all \(m!\) wire mapping schemes (for \(m\) split qubits). Our inference-based attack challenges this assumption: we demonstrate that only a small number of I/O queries, combined with the techniques described herein, are sufficient to reconstruct hidden connections across splits.
Specifically, we use a hierarchical pruning approach that separates Split 2 into multiple qubit blocks, each with a non-overlapping set of qubits. 
This reduces the total complexity to $\sum_{b=1}^m n_b b!$, where $\sum_{b=1}^m n_bb=m$, $n_b$ is the number of qubit blocks with $b$ qubits, 
which significantly reduces the computational load when not all qubits entangle in Split 2. Practice demonstrates that rapid convergence is achieved by eliminating most candidate schemes with only a few queries during the single-wire or paired-wire stages.


 


\section{Experiments and Results}
\subsection{Security Evaluation Protocol and Metrics}
The goal of our work is to evaluate the security levels of current split compilation techniques for quantum circuit obfuscation. To this end, it is essential to measure the impact of splitting location choice on the attack complexity in order to provide constructive feedback to the quantum circuit designers.
In our experiments, we increment the depth (number of layers) in Split 2, denoted by $n$, from 1 to the entire circuit but the first layer.
We record the attack complexity in terms of the number of oracle queries $t$ for each possible value of $n$.
In this way, we can observe the trend and enable designers to predict the attack complexity at design time.

\subsection{Experiment Setup}
We conducted experiments using the IBM Qiskit framework to compile and simulate quantum circuits. We employ the RevLib benchmark suite \cite{wille2008revlib} which contains quantum circuits extensively utilized in quantum computing research. These benchmark circuits comprise a diverse set of gates, and their total depths range from 4 to 29 layers, with qubit sizes varying between 4 and 15 qubits. 
For each circuit, we test how the number of layers in Split 2 $n$ affect the number of matches by varying . 
If the original circuit depth is $L$, then Split 2 contains layers $L-n+1,\dots,L$, while Split 1 contains layers $1,\dots,L-n$. 
We implemented the proposed connection-recovery framework in IBM Qiskit and ran all experiments on the Qiskit Aer simulator. 
We use the FakeValencia backend and incorporate noise into the simulation. We choose error threshold $\epsilon=0.03$.


\subsection{Result Analysis}
\begin{figure}[htb]
  \centering
  \includegraphics[width=\linewidth]{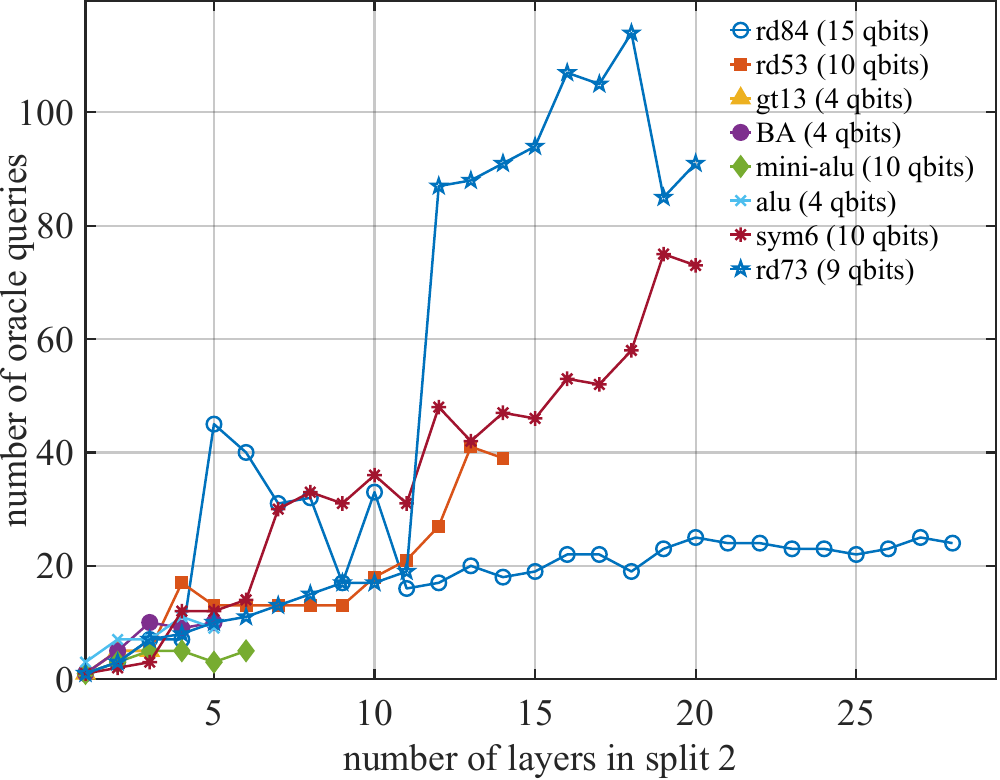} 
  \caption{The x-axis displays the number of layers in Split 2 ($n$);
  the y-axis displays the number of matching attempts $t$.}
  \label{fig:overview_all}
\end{figure}

Fig. \ref{fig:overview_all} compares the variation in the number of attempts required for matching from oracle queries across multiple benchmark circuits at different Split 2 layer depths. The horizontal axis represents the number $n$ of Split 2 layers, while the vertical axis shows the required number $t$ of oracle queries. Each line represents a benchmark (legend includes circuit size: rd84 is 15 qubits, rd73 is 9 qubits, rd53 is 10 qubits, sym6 is 10 qubits, mini-alu is 10 qubits, gt13 / BA / alu is 4 qubits).
Except for a few inflection points, most benchmarks exhibit an increase in query counts as the number of layers n grows, with more layers increasing the matching difficulty. 
This is expected because, as more gates are incorporated into Split 2, more qubits tend to entangle together, which increases the attack complexity.
We can also observe from the figure that when Split 1 has fewer layers, the attack complexity decreases to some extent.
This is because when Split 1 has very few layers, some qubits in Split 1 may not have any gates, allowing the attacker to control the input to Split 2 directly and thus reducing attack complexity. For a quantum circuit designer, this means that the optimal way to split a quantum circuit is to have just a few layers in Split one and the rest of the circuit in Split 2.

Although the attack complexity generally increases with the depth of Split 2, $n$, we can see that the growth is mostly linear, and the maximum complexity is still very low. This is in stark contrast with the claimed attack complexity, i.e., exponential in the number of qubits, in prior work \cite{saki2021split, wang2025tetrislock}. Our results suggest that simply splitting the quantum circuit in two parts is not secure against oracle-guided attacks.


\section{Conclusion}
We propose a framework that challenges the widely held assumption that split compilation can reduce attackers' capabilities to only brute-force search attacks. In this work, we take the 2-split compilation as example and decompose Split 2 into disjoint blocks using its qubit entanglement pattern. This allows us to reconnect both splits with minimal computational effort.A fully leveraging the reversibility and structural regularity inherent in the splitting process.
Among all tested circuits, the number of oracle queries $t$ required for recovery mapping generally increases with Split 2 depth $n$, yet remains significantly lower than the exponential complexity claimed in existing work. 

This study assumes known oracle I/O and qubit order in Split 1; experiments are conducted using the Qiskit simulator. Future work will relax these assumptions by addressing partial knowledge and multi-split scenarios; derive information-theoretic lower bounds on query complexity; and co-design compiler heuristics explicitly targeting attack-aware splitting. We expect this evaluation paradigm to advance the establishment of a more rigorous methodology for attack-aware quantum intellectual property protection.


\section*{Acknowledgment}
This work is supported by the National Science Foundation under Award 2530705.

\balance

\bibliographystyle{IEEEtran}
\bibliography{qref}

\end{document}